\documentclass[%
reprint,
nofootinbib,
amsmath,amssymb,
aps,
]{revtex4-2}

\usepackage{graphicx}
\usepackage{amssymb}
\usepackage{amsmath}
\usepackage{mathbbol}
\usepackage{amsthm}
\usepackage[normalem]{ulem}
\usepackage{slashed}
\usepackage{color,rotating}
\usepackage{bbm}
\usepackage{array}
\usepackage[utf8]{inputenc}
\usepackage{multirow}
\usepackage{xcolor}
\usepackage{leftidx}
\usepackage[normalem]{ulem}

\allowdisplaybreaks
\usepackage[toc,page]{appendix}

\newcommand{\p}{\partial}

\newcommand{\cG}{\mathcal{G}}
\newcommand{\cM}{\mathcal{M}}

\newcommand{\cR}{\mathcal{R}}

\newcommand{\be}{\begin{equation}}
\newcommand{\ee}{\end{equation}}

\begin{document}

\title{Foliated Asymptotically Safe Gravity \\ Lorentzian
Signature Fluctuations from the Wick Rotation}

\author{Frank Saueressig}
\email{f.saueressig@science.ru.nl}

	\affiliation{Institute for Mathematics, Astrophysics and Particle Physics (IMAPP) \\ Radboud University, Heyendaalseweg 135, 6525 AJ Nijmegen,The Netherlands 
}

\author{Jian Wang}
\email{{jianwang2024@ucas.ac.cn}}

	\affiliation{Institute for Mathematics, Astrophysics and Particle Physics (IMAPP) \\ Radboud University, Heyendaalseweg 135, 6525 AJ Nijmegen,The Netherlands 
}
\affiliation{School of Fundamental Physics and Mathematical Sciences\\
Hangzhou Institute for Advanced Study, UCAS, Hangzhou 310024, China}

\begin{abstract}
Asymptotic Safety constitutes a promising mechanism for a consistent and predictive high-energy completion of the gravitational interactions. To date, most results on the interacting renormalization group fixed point underlying the construction are obtained for Euclidean signature spacetimes. In this work, we use the Arnowitt-Deser-Misner (ADM) decomposition of the metric degrees of freedom and investigate the relations between the Euclidean and Lorentzian renormalization group flows resulting from the analytic continuation of the lapse function. We discuss the general conditions which guarantee the equivalence of the beta functions. These insights are illustrated based on the flow of the graviton two-point function within the Einstein-Hilbert truncation, demonstrating agreement of the Euclidean and Lorentzian settings. Hence the UV- and IR-completions identified in the Euclidean case are robust when changing spacetime signature. We take this as an important indicator that the Euclidean asymptotic safety mechanism carries over to Lorentzian signature spacetimes. 
\end{abstract}

\maketitle

%\begin{keyword}
%Functional Renormalization Group \sep Asymptotic Safety \sep Gravity-Matter Systems \sep 

%\PACS{04.60.-m,04.62.+v,11.10.Hi} 

%\end{keyword}

%\end{frontmatter}

%-------------------------------------------------------
%\include{ADM}
%\include{FRGE}
%\include{RG flow pure gravity}
%-------------------------------------------------------
\section{Introduction}
%-------------------------------------------------------
The construction of a ultraviolet (UV)-complete quantum field theory of gravity is still an open question in modern theoretical physics. One candidate for such a theory is the gravitational asymptotic safety program \cite{Percacci:2017fkn,Reuter:2019byg} also reviewed in \cite{Saueressig:2023irs,Pawlowski:2023gym,Knorr:2022dsx,Morris:2022btf,Martini:2022sll,Wetterich:2022ncl,Codello:2008vh,Nagy:2012ef,Reichert:2020mja}. Starting from Weinberg’s initial conjecture \cite{Weinberg:1980gg}, the Wetterich equation \cite{Wetterich:1992yh,Morris1994} adapted to gravity \cite{Reuter:1996cp}, has provided substantial evidence that the gravitational renormalization group (RG) flow indeed possesses an interacting fixed point which could render gravity asymptotically safe \cite{Christiansen:2017bsy,Denz:2016qks,Christiansen:2015rva,Christiansen:2014raa,Eichhorn:2015bna, Gies:2022ikv,Gies:2016con,deBrito:2020xhy,DeBrito:2018hur,Kluth:2022vnq}. This fixed point, called the Reuter fixed point, also extends to a variety of gravity-matter systems \cite{Pastor-Gutierrez:2022nki,Burger:2019upn,Pawlowski:2018ixd,Eichhorn:2018ydy,Eichhorn:2018akn,Christiansen:2017cxa,Meibohm:2016mkp,Meibohm:2015twa,deBrito:2022vbr,Eichhorn:2021qet,deBrito:2021pyi,deBrito:2020dta,Eichhorn:2019ybe,Eichhorn:2018nda,Eichhorn:2018whv,Eichhorn:2017als,Eichhorn:2017ylw,Dona:2015tnf,Knorr:2022ilz,DeBrito:2019rrh}, making the program attractive for a wide range of phenomenological applications including particle physics \cite{Eichhorn:2022gku}, black hole physics \cite{Eichhorn:2022bgu,Platania:2023srt}, and cosmological applications \cite{Bonanno:2017pkg,Bonanno:2024xne}. Links to other approaches to quantum gravity, including Causal Dynamical Triangulations (CDT) \cite{Ambjorn:2024qoe, Ambjorn:2024bud}, canonical quantization \cite{Thiemann:2024vjx,Ferrero:2024rvi}, and swampland conjectures \cite{Basile:2021krr,Knorr:2024yiu,Eichhorn:2024wba,Eichhorn:2024rkc} have recently been explored as well.

Most investigations related to Asymptotic Safety are carried out in a Euclidean background spacetime. In order to arrive at a more realistic description of nature spacetime should come with Lorentzian signature though. Drawing inspiration from quantum field theory in a curved spacetime \cite{Birrell:1982ix}, this transition introduces a series of new elements
\begin{enumerate}
\item As compared to the Euclidean setting, the Lorentzian setup requires a new geometric structure - essentially a preferred direction - which is associated with time. This affects, e.g., the space of fluctuations in the metric field \cite{Demmel:2015zfa}.
\item In contrast to the Euclidean case, propagators and the choice of vacuum are no longer unique as soon as the background spacetime is sufficiently generic.
\item In a quantum field theory defined on a flat, non-dynamical Euclidean background, the analytic continuation from a Euclidean to Lorentzian spacetime is implemented by the Wick rotation of the time-coordinate,  $\tau^{\rm E} \rightarrow i \tau^{\rm L}$. When it comes to gravity on a general curved spacetime, this prescription could lead to complex and unphysical metrics if $g_{\mu\nu}$ depends on time explicitly \cite{Baldazzi:2018mtl}. This may be bypassed by an analytic continuation of the lapse function, as studied in detail in the first half of \cite{Banerjee:2024tap}.
\end{enumerate}
These points posit a clear mandate for investigating the working of the asymptotic safety mechanism also for Lorentzian signature spacetime, leading to the Lorentzian asymptotic safety program.

Starting from \cite{Manrique:2011jc}, the impact of spacetime signature has been explored from various angles. Formally, one can derive the Wetterich equation from a path integral formulated in Lorentzian signature. Borrowing tools developed in the context of algebraic quantum field theory allows to formulate this equation in a background-independent way \cite{DAngelo:2022vsh,DAngelo:2023tis}. This strategy highlights the state-dependence of the Lorentzian construction \cite{DAngelo:2023wje,Banerjee:2022xvi}. On the technical side, these developments have been complemented by developing heat-kernel methods for Lorentzian metrics on real manifolds  \cite{Banerjee:2024tap}. Along a different path, the fluctuation approach towards solving the Wetterich equation \cite{Pawlowski:2023gym} has started from a flat Minkowski background and constructed the spectral function of the graviton on this background \cite{Bonanno:2021squ,Fehre:2021eob}. Based on these results, it was argued that the spectral function comprises a massless one-graviton peak and a multi-graviton continuum with an asymptotically safe scaling for large momenta.

The points discussed above lead to the intriguing question whether there is a canonical relation among RG flows obtained in the Euclidean and Lorentzian settings. From a geometrical perspective, the first step towards such a connection is to equip the Euclidean spacetime with a foliation structure.\footnote{Such a foliation structure is also essential when defining Causal Dynamical Triangulations \cite{Ambjorn:2012jv,Loll:2019rdj,Ambjorn:2024pyv}. It also provides the key element in Ho\v{r}ava-Lifshitz gravity \cite{Horava:2009uw} where an anisotropy between space and time is used to arrive at a perturbatively renormalizable quantum field theory, see \cite{Barvinsky:2015kil,Barvinsky:2021ubv,Barvinsky:2023uir} for recent results.} This can either be done by applying the Arnowitt-Deser-Misner (ADM) decomposition to the metric \cite{Arnowitt:1962hi,Gourgoulhon:2007ue} or adding additional geometric objects to the covariant formulation \cite{Knorr:2018fdu}. Asymptotic Safety based on the ADM formalism has already been investigated in a series of works within the background approximation \cite{Manrique:2011jc,Biemans:2016rvp,Biemans:2017zca} and also in the fluctuation approach focusing on the graviton propagator \cite{Saueressig:2023tfy,Korver:2024sam}. A remarkable insight obtained from this line is that the ADM and covariant formulations give rise to very similar results with respect to the existence of interacting renormalization group fixed points and phase diagrams. This is non-trivial, since the two constructions encode the fluctuating degrees of freedom in different fields so that one should expect to deal with different quantum theories on generic grounds \cite{Wetterich:2024ivi}.

In this work, we complement our understanding of RG flows within the ADM formalisms by studying the Wick rotation from Euclidean to Lorentzian signature spacetimes through the analytic continuation of the lapse function. This avoids the issues related to complex metrics which arises from the analytic continuation of the time coordinate. Furthermore, this prescription may restore the conventional causal Feynman propagator on a Minkowski spacetime \cite{Baldazzi:2018mtl}. We then identify generic conditions on the propagators and regulators entering in the evaluation of the Wetterich equation which ensure that RG flows in the Euclidean and Lorentzian settings agree.

As a concrete application, we compute the RG flow of the graviton two-point function using the interaction vertices generated by the Einstein-Hilbert action and show that the resulting beta functions obtained in the Lorentzian case are identical to the Euclidean ones.  Based on the RG flow, we then identify a suitable interacting fixed point with two relevant directions. This suggests that also Lorentzian quantum gravity can be UV-complete via the asymptotic safety mechanism and constitutes a key step towards relating the Lorentzian asymptotic safety program to existing Euclidean results.

This paper is organized as follows. In Section \ref{sect.2}, we introduce the ADM decomposition and review the prescription of Wick rotation via complexifying the lapse function. Section \ref{sect.3} reviews the Wetterich equation and specifies the setup underlying our computation. Our results for the beta functions are given in Section \ref{sect.4} and the resulting fixed-point structure and phase diagrams are constructed in Section \ref{sect.4b}. We close with a discussion of our results and outlook in Section \ref{sect.5}. The explicit form of the ghost sector entering our computation is given in Appendix \ref{App.B} and we report on the relations between the fixed points identified in this and previous works in Appendix \ref{App.C}.

%-------------------------------------------------------
\section{Foliated spacetime \newline and Wick rotation}
\label{sect.2}
%-------------------------------------------------------
We start by reviewing the ADM formalism \cite{Arnowitt:1962hi,Gourgoulhon:2007ue}. We take the spacetime $\cM$ as a $d+1$-dimensional manifold admitting a foliation. This spacetime is equipped with coordinates $x^\mu$, $\mu=\{ 0,\cdot \cdot \cdot, d\}$ and a metric $g_{\mu\nu}$. Subsequently, we introduce a scalar function $\tau(x)$. This function foliates $\cM$ into a one-parameter family of hypersurfaces $\Sigma_\tau$, containing all points with the same value for $\tau$. The existence of such a foliation is a necessary ingredient for transiting from Euclidean to Lorentzian signature spacetimes and constitutes a geometrical structure which is not necessarily present in the Euclidean signature case.

The foliation structure allows us to introduce a new coordinate system, given by the time-coordinate $\tau$ and coordinates $y^i$, $i=1, \cdots, d$ providing a coordinate system on $\Sigma_\tau$. Assuming Euclidean signature, the line element written in terms of the coordinates $\{\tau, y^i\}$ can be given as 
\begin{equation}\label{LineEle}
ds^2 = N^2 d\tau^2 +\sigma_{ij} (N^i d\tau + dy^i) (N^j d\tau +dy^j).
\end{equation}
This decomposition encodes the metric degrees of freedom in the lapse function $N$, the shift vector $N^i$, and the metric $\sigma_{ij}$ measuring distances on $\Sigma_\tau$. From Eq.\ \eqref{LineEle}, the components of the metric tensor are read off as 
\begin{equation}\label{metreucadm}
g^{}_{\mu\nu }=
\begin{pmatrix}
N^2+N^i N_i  ~~~& N_j\\
N_i~~~ & \sigma_{ij}\\
\end{pmatrix}.
\end{equation}

Now, we consider the transition from Euclidean to Lorentzian signature. One way to implement this on a generic, curved spacetime is the analytic continuation of the lapse function along a curve in the complex plane \cite{Baldazzi:2018mtl,Banerjee:2024tap},
\begin{equation}\label{redeflap}
N  \mapsto \sqrt{\epsilon_c} \,  N \, .
\end{equation}
This generalizes the Wick flip where $\epsilon_c = \pm 1$ takes discrete values \cite{Dasgupta:2001ue,Rechenberger:2012dt}. In the generalization \eqref{redeflap}, the parameter $\epsilon_c$ encodes the contour $c$ taken in the analytic continuation. Substituting this expression into Eq.\ \eqref{metreucadm} leads to
 \begin{equation}\label{metrepsadm}
g^{(\epsilon_c)}_{\mu\nu }=
\begin{pmatrix}
\epsilon_c N^2+N^i N_i  ~~~& N_j\\
N_i~~~ & \sigma_{ij}\\
\end{pmatrix}.
\end{equation}
One can easily verify that this parametrized metric interpolates between the Euclidean metric for $\epsilon_c = +1$ and the Lorentzian metric for $\epsilon_c = -1$.
As pointed out by \cite{Baldazzi:2018mtl}, if one treats the parameter $\epsilon_c$ as a continuous real number, the inverse of the metric \eqref{metrepsadm} will degenerate at $\epsilon_c=0$. This suggests to regard $\epsilon_c$ as complex with the analytic continuation either along the unit circle \cite{Banerjee:2024tap} in the complex plane or by shifting the path along the real line by a small imaginary part \cite{Baldazzi:2018mtl}. We will consider the second option, setting
\begin{equation}\label{paramepsil}
\epsilon_c =\epsilon_s + i \, c \, \epsilon \, ,
\end{equation}
with $c$ being a real, positive, and field-dependent normalization factor. The real part $\epsilon_s$ takes values in the interval $[-1,1]$. The lower and upper boundary corresponds to the Lorentzian and Euclidean signature metric, respectively. The parameter $\epsilon$ ensures that the metric remains regular along the integration contour and is treated as infinitesimal.

It is instructive to illustrate the effect of this analytic continuation at the level of a free scalar field $\phi$ of mass $m$. Setting $N_i = 0$ for clarity, the ADM-decomposed action is
\be\label{scalaractionADM}
\begin{split}
& S[\phi] =  \frac{1}{2} \int d\tau d^dy \, \sqrt{\epsilon_s} N \sqrt{\sigma} \times \\ 
& \, \, \left( \frac{1}{\epsilon_s N^2} (\p_\tau \phi)^2 + \sigma^{ij} (\p_i \phi) (\p_j \phi) + m^2 \phi^2 - i \epsilon \phi^2\right).
\end{split}
\ee
Here we expanded to first order in $\epsilon$ and chose the constant $c$ in \eqref{paramepsil} to ensure standard normalization. Moreover, we dropped the $i\epsilon$-terms coming from the expansion of the determinant of the spacetime metric, $\sqrt{\epsilon_c} \, N\sqrt{\sigma}$-factor, since they play no role in fixing the analytic structure of the two-point function.\footnote{Taking into account these terms leads to two-point functions that are qualitatively similar (but not identical) to the one considered by Zimmermann \cite{Zimmermann:1968mu}, also see \cite{Duncan:2012aja} for a pedagogical account. While, ultimately, this may aid in the convergence of loop diagrams in the Lorentzian setting, this is not relevant for us in the sequel.}

Based on \eqref{scalaractionADM}, it is straightforward to show that one recovers the causal Feynman propagator on a flat Lorentzian spacetime by inverting the second functional derivative of $S[\phi]$. Setting $N=1$ and $\sigma^{ij} = \delta^{ij}$ and converting to momentum space, denoting the time- and spatial components of the momentum four-vector by $p_0$ and $\vec{p}$, respectively, yields
\be\label{propscalar}
\cG = \frac{1}{ \sqrt{\epsilon_s} \left( \epsilon_s^{-1} p_0^2 + \vec{p}^{~2} + m^2 - i \epsilon \right)} \, . 
\ee
In the Euclidean signature case where $\epsilon_s = 1$ the denominator is positive definite and the $i \epsilon$-term can be dropped. In the Lorentzian case, Eq.\ \eqref{propscalar} evaluates to
\be\label{propscalar1}
\cG = \frac{i}{  \left(  p_0^2 - \vec{p}^{~2} - m^2 + i \epsilon \right)} \, , 
\ee
and one recovers the standard $i\epsilon$ prescription of a massive scalar propagator in mostly-plus signature. Hence the prescription \eqref{paramepsil} recovers the standard Wick-rotation results in a flat spacetime. At the same time, it also applies to the more general case where spacetime is curved, thereby giving a unique relation between Euclidean and Lorentzian signature results. In the remaining part of this paper, we will apply this analytic continuation to the Wetterich equation adapted to the ADM formalism \cite{Manrique:2011jc,Rechenberger:2012dt}, explicitly relating RG flows obtained for Euclidean and Lorentzian signature backgrounds.

%-------------------------------------------------------
\section{Wetterich equation and \newline projection schemes}
\label{sect.3}
%-------------------------------------------------------
Investigating the asymptotic safety mechanism in the context of gravity requires a method to compute RG flows. The main tool used in these studies is the Wetterich equation \cite{Wetterich:1992yh,Morris1994},
\begin{equation}\label{WettEq}
k \partial_k \Gamma_k = \frac{1}{2} \text{STr} \left[ \cG_k \, k \partial_k R_k \right].
\end{equation} 
 The Wetterich equation encodes the dependence of the effective average action $\Gamma_k$ on the coarse-graining scale $k$. Its right-hand side (rhs) contains the scale-dependent propagator $\cG_k \equiv \left( \Gamma_k^{(2)} + R_k \right)^{-1}$ where $\Gamma_k^{(2)}$ is the second functional derivative of $\Gamma_k$ with respect to the quantum field. The propagator has been supplied by a regulator $R_k$ which provides a mass term to fluctuations with momentum $p^2 \lesssim k^2$ and vanishes for $p^2 \gg k^2$. The supertrace STr contains the integral over the internal momenta, a sum over the fields, and depends on the signature of spacetime through $\epsilon_s$. The interplay of $R_k$ in the propagator and the numerator ensures that the trace contribution is finite and peaked on momenta $p^2 \approx k^2$. In this way, the Wetterich equation realizes Wilson's idea of renormalization, integrating out quantum fluctuations shell-by-shell in momentum space when lowering the coarse-graining scale. For convenience, we will trade $k$ for the dimensionless RG time $t \equiv \text{ln}(k/k_0)$ where $k_0$ is an arbitrary reference scale.

The goal of this work is to use Eq.\ \eqref{WettEq} to investigate the signature-dependence of the gravitational RG flow. For this purpose, we encode the gravitational degrees of freedom in the ADM fields so that the Euclidean and Lorentzian settings can be connected by the analytic continuation of the lapse function \eqref{redeflap}. We then employ the background field method, splitting the ADM fields into background fields (conventionally distinguished by a bar) and fluctuations (coming with a hat)
\begin{equation}\label{linsplit}
\begin{split}
\sigma_{ij}=\bar{\sigma}_{ij}+\hat{\sigma}_{ij}, \,\,\, N_i =\bar{N}_i +\hat{N}_i,\,\, \, N=\bar{N}+\hat{N}.
\end{split}
\end{equation}
For the present study, it is sufficient to adopt a flat background with $\bar{N}=1$, $\bar{N}_i=0$, and $\bar{\sigma}_{ij}=\delta_{ij}$. Correlation functions of the fluctuation fields can then be obtained by taking the functional derivatives of $\Gamma_k$ with respect to the fluctuations.\footnote{Conceptually, one should distinguish between the cases where the analytic continuation from Euclidean to Lorentzian signature is implemented at the level of the lapse function $N$ as in Eq.\ \eqref{redeflap} or the background lapse function via $\bar{N} \mapsto \sqrt{\epsilon_c} \, \bar{N}$. We choose to work with the former. The projective property of the Wetterich equation \eqref{WettEq}, ensuring that any constant rescaling a field drops out of the equation, in combination with the linear split \eqref{linsplit} guarantees that both choices give the same result.}

In order to reduce the complexity of the computation, we apply the York decomposition \cite{York1973} to the fluctuation fields
\begin{equation}
	\label{eqn: York decomp metric}
	\hat{\sigma}_{ij}=h_{ij}+\partial_i \frac{1}{\sqrt{\bar{\Delta}}}v_j+\partial_j \frac{1}{\sqrt{\bar{\Delta}}}v_i +\partial_i \partial_j \frac{1}{\bar{\Delta}}E+\frac{1}{3}\delta_{ij} E+\frac{1}{3}\delta_{ij}\psi \, . 
\end{equation}
Here the background Laplacian is defined as $\bar{\Delta} \equiv - \delta^{ij} \partial_i \partial_j$, and the component fields satisfy
\begin{equation}
	\partial^i h_{ij}=0, \quad \delta^{ij}h_{ij}=0, \quad \partial^i v_i=0, \quad\psi =\delta^{ij}\hat{\sigma}_{ij} \, . 
\end{equation}
 The decomposition for the shift vector is
\begin{equation}
	\label{eqn: York decomp shift}
	\hat{N_i}=u_i+\partial_i \frac{1}{\sqrt{\bar{\Delta}}}B, \quad \partial^i u_i=0 \, . 
\end{equation}
Note that this decomposition is independent of the lapse function $N$. Hence it is unaffected by the analytic continuation \eqref{redeflap}.

In order to be able to compute the RG flow explicitly we then approximate $\Gamma_k$
 by the Einstein-Hilbert (EH) action supplemented by gauge-fixing (gf) and ghost terms,
\begin{equation}\label{actionansatz}
\Gamma_k \simeq \Gamma^{\text{EH}}_k + \Gamma^{\text{gf}}_k+ \Gamma^{\text{ghost}}.
\end{equation}
The Einstein-Hilbert action written in terms of the ADM fields and the substitution \eqref{redeflap} is given by
\begin{equation}\label{EH-ADM}
\begin{split}
\Gamma^{\text{EH}}_k =  & \frac{1}{16 \pi G_k} \int d\tau d^{3} y  \, \sqrt{\epsilon_s}\, N \sqrt{\sigma} \times \, \\ & \left(  \epsilon^{-1}_s K^{ij}K_{ij} - \epsilon^{-1}_s K^2-R + 2\Lambda_k + i \epsilon \right) \,.
\end{split}
\end{equation}
Here the extrinsic curvature $K_{ij}$ is defined as
\begin{equation}
K_{ij} \equiv \frac{1}{2 N} \left( \partial_\tau \sigma_{ij} -D_i N_j -D_j N_i\right),
\end{equation}
with $D_i$ being the covariant derivative defined with respect to the spatial metric $\sigma_{ij}$, $R$ is the Ricci scalar constructed from $\sigma_{ij}$, and the $i\epsilon$ acts as a reminder that for Lorentzian signature poles in the propagator are shifted away from the axis of integration. The approximation \eqref{EH-ADM} tracks the flow of two couplings, Newton's coupling $G_k$ and the cosmological constant $\Lambda_k$ which have been promoted to functions of the coarse-graining scale $k$.

Finally, we need to specify gauge-fixing and ghost terms. As pointed out by \cite{Biemans:2016rvp,Houthoff:2017oam}, the two-point vertices generated by the Einstein-Hilbert action in the ADM formalism are non-relativistic in general. This can be fixed by adopting harmonic gauge \cite{Biemans:2017zca}. Including the terms quadratic in the fluctuations then gives the following gauge-fixing action,
\begin{equation}
 	\label{eqn:gfaction}
 	\Gamma^{\text{gf}}_k=\frac{1}{32 \pi G_k}\int \mathrm{d}\tau \mathrm{d}^3 y  \sqrt{\epsilon_s} \left(\epsilon_s F^2 +\delta^{ij}F_{i} F_{j}\right) \, , 
 \end{equation}
with the gauge-fixing conditions,
\begin{equation}\label{eqn:gffunctionals}
 \begin{split}
 	F &= \frac{1}{ {\epsilon_s}}\partial_{\tau}\hat{N}+ \frac{1}{ {\epsilon_s}} \partial^i \hat{N}_i - \frac{1}{ {\epsilon_s}}\frac{1}{2} \partial_{\tau} \psi \, , \\
 	F_i &=  \frac{1}{ {\epsilon_s}} \partial_{\tau}\hat{N}_i - \partial_i \hat{N} - \frac{1}{2} \partial_i  \psi +  \partial^j \hat{\sigma}_{ij} . 
 \end{split}
 \end{equation}
The ghost term resulting from this choice is obtained by the standard Faddeev-Popov procedure and is given in Appendix \ref{App.B}.

According to \cite{Bardeen:1980kt,Craps:2014wga}, the transverse-traceless (TT)-mode $h_{ij}$ encodes the tensor fluctuations in a gauge-invariant way. It is the two-point function of this field which is related to observations. Therefore, we will read off the flow equations for $G_k$ and $\Lambda_k$ from the propagator of $h_{ij}$. Conceptually, this corresponds to a fluctuation field computation along the lines described in \cite{Pawlowski:2020qer,Pawlowski:2023gym}. In general, the scale dependence of an $n$-point vertex is obtained from \eqref{WettEq} by taking derivatives with respect to the corresponding fluctuation fields, collectively denoted by $\hat{\chi}$. Schematically,
\begin{equation}\label{WettEqForGamN}
\partial_t \Gamma^{(n)}_k = \frac{1}{2} \text{STr} \left[\frac{\delta^n}{\delta \hat{\chi}^n} \, \cG_k \,  \partial_t R_k \, \right].
\end{equation} 
For the two-point correlation function of $h_{ij}$ this general expression evaluates to
\begin{equation}\label{flowtwopoint}
\begin{split}
\partial_t \Gamma^{(hh)}_k = & \; {\rm STr}\Big[ \cG_k  \; \Gamma_k^{(3)} \; \cG_k   \Gamma_k^{(3)} \; \cG_k \;  \p_t \cR_k \Big] \\
& - \frac{1}{2}{\rm STr}\left[ \cG_k  \; \Gamma_k^{(4)} \; \cG_k  \; \p_t \cR_k \right] \, .
\end{split}
\end{equation}
Here $\Gamma_k^{(3)}$ and $\Gamma_k^{(4)}$ denote the three- and four-point vertices obtained from $\Gamma_k$. Here, it is tacitly understood that the terms on the rhs come with two external graviton legs and momentum $p^2$. In order to close the system, we then generate the interaction vertices from \eqref{actionansatz}. This completes our computational framework.

The first step in evaluating \eqref{flowtwopoint} is the construction of the required $n$-point vertices. This can be easily done using computer algebra software \cite{xactref,Nutma:2013zea}. The two-point functions resulting from this procedure are collected in Table \ref{table.hessian.parame}.
\begin{table}[t!]
	\renewcommand{\arraystretch}{1.5}
	\begin{center}
		\begin{tabular}{ll}
			\hline \hline
			fields \hspace*{4mm} & components of $\Gamma_k^{(2)}$ \\ \hline \hline
			$h_{ij} h^{kl}$ & $\frac{1}{32 \pi G_k}\sqrt{\epsilon_s} \left(  ({\epsilon_s^{-1}} p_0^2 + \vec{p}^{\,2})-2 \Lambda_k \right) \, {\Pi_h}^{ij}_{~~kl} $ \\ \hline 
			$v_i v^j$ & 
			$\frac{1}{16 \pi G_k}\sqrt{\epsilon_s}  \left( ({\epsilon_s^{-1}} p_0^2 + \vec{p}^{\,2})- 2 \Lambda_k \right) {\Pi_u}{}^i{}_j$ \\
			$EE$ & 
			$\frac{1}{48 \pi G_k} \sqrt{\epsilon_s} \left(  ({\epsilon_s^{-1}} p_0^2 + \vec{p}^{\,2}) - 2 \Lambda_k \right)$ \\
			$\Psi\Psi$ & 
			$- \frac{1}{192 \pi G_k} \sqrt{\epsilon_s}  \left({\epsilon_s^{-1}}   p_0^2 +\vec{p}^{\,2} -  2 \Lambda_k \right)$ \\
			$\hat{N}\hat{N}$ & 
			$\frac{1}{16 \pi G_k}\sqrt{\epsilon_s}  ({\epsilon_s^{-1}} p_0^2+\vec{p}^{\,2})$ 	\\ 
			$\Psi \hat{N}$ & 
			$-\frac{1}{16 \pi G_k}  \sqrt{\epsilon_s} \left({\epsilon_s^{-1}}   p_0^2+ \vec{p}^{\,2} -2 \Lambda_k \right)$ \\
			$u^i u_j$ & 
			$\frac{1}{16 \pi G_k} {\epsilon_s^{-1/2}}({\epsilon_s^{-1}} p_0^2+ \vec{p}^{\,2}) {\Pi_u}{}^i{}_j $ 	\\ 
			$BB$ & 
			$\frac{1}{16 \pi G_k}{\epsilon_s^{-1/2}} \left({\epsilon_s^{-1}} p_0^2+ \vec{p}^{\,2} \right)$ \\ \hline
			$\bar{c}c$ & $- \, \sqrt{\epsilon_s}  (p^2_0 +\epsilon_s  \vec{p}^{\,2})$ \\
			$\bar{b}^i b_i$ & $- \,\sqrt{\epsilon_s} ({\epsilon_s^{-1}}  p^2_0 +\vec{p}^{\,2}) \, \Pi_u{}_j{}^i$
			\\ \hline \hline
		\end{tabular}
	\end{center}
	\caption{\label{table.hessian.parame} Matrix elements of the two-point functions $\Gamma^{(2)}_k$ including the analytic continuation of the lapse function \eqref{paramepsil}. Here $\Pi$ are the standard tensor structures associated with the internal indices of the fluctuation fields \cite{Percacci:2017fkn} and the $i\epsilon$ terms relevant in the Lorentzian setting are suppressed for the sake of readability.}
\end{table}
Owed to the gauge choice \eqref{eqn:gfaction}, all two-point functions (and propagators) come with a relativistic dispersion relation. Moreover, the cosmological constant serves as a mass term in some of the two-point functions. This also includes the correlation function for $h$.

Next, we need to specify the regulator $R_k$. In the Euclidean setting, the quantity separating the fluctuations into high- and low-momentum modes is taken as the magnitude of the momentum four-vector $p^2 = p_0^2 + \vec{p}^2$ which is positive definite. This leads to an intuitive ordering in which fluctuations with large four-momentum are integrated out first. The Lorentzian signature analogue $p^2 = - p_0^2 + \vec{p}^2$ is no longer positive definite though, making the discrimination of high- and low-momentum modes more subtle. In addition, the regulator may alter the analytic properties of $\cG_k$, by inducing new poles in the complex momentum plane. Such an improper regularization procedure may then spoil the connection between the evaluation of the trace in \eqref{WettEq} in Euclidean and Lorentzian signature backgrounds, also see \cite{Braun:2022mgx} for a related discussion.

In order to ameliorate these subtleties, we opt for a regulator which discriminates high- and low-momentum modes according to their spatial momentum $\vec{p}^{\,2}$. Based on this choice, we follow \cite{Manrique:2011jc,Banerjee:2022xvi} and implement the Type I regularization by replacing
\begin{equation}\label{Rchoice}
\vec{p}^{\,2} \mapsto \vec{p}^{\,2} + R_k(\vec{p}^{\,2}).
\end{equation}
The regulator function $R_k(\vec{p}^{\,2})$ is taken of Litim-type 
\be\label{reg-spatial}
R_k(\vec{p}^{\,2}) = (k^2-\vec{p}^{\,2}) \Theta(k^2-\vec{p}^{\,2}) \, ,
\ee
with $\Theta$ being the Heaviside step function \cite{Litim:2000ci,Litim:2001up}. This is sufficient to render the momentum integrals in the STr finite, thus meeting the finiteness criterion (finiteness). It also guarantees that there are no additional poles appearing in the complex $p_0$-plane as this choice also acts as a mass term in the Lorentzian signature setting (analyticity). These features come at the expense that the regulator breaks Lorentz invariance explicitly though. Thus, the structural features of the choice \eqref{Rchoice} are summarized as follows
\be
\begin{array}{c|c|c}
 & \quad {\rm yes \quad}  & \quad {\rm no \quad}  \\ \hline
\Big. {\rm finiteness} & {\bf X} & \\ 
\Big.  {\rm analyticity} & {\bf X} & \\ 
\Big.  \quad {\rm Lorentz \; symmetry \quad}  & & {\bf X} \\ \hline
\end{array}
\ee
Currently, there is no known regularization function which meets all three criteria simultaneously. It is the trade between analyticity and Lorentz symmetry which then distinguishes the present analysis from the previous work \cite{Saueressig:2023tfy} and also the regularization procedures employed in the fluctuation computations in the covariant setting \cite{Pawlowski:2023gym}.

Finally, we specify our projection scheme. Our goal is to determine the flow of the two-point function $\Gamma_k^{(hh)}$ given in the first line of Table \ref{table.hessian.parame}. Hence we project onto
\be\label{projection-space}
\Gamma^{(hh)}_k = \frac{1}{32 \pi G_k}\sqrt{\epsilon_s} \left( (\epsilon^{-1}_s p^2_0 + \vec{p}^{\, 2}) -2 \Lambda_k \right) {\Pi_h}^{ij}_{kl} \, .
\ee
In this way $-2 \Lambda_k \equiv \mu_k^2$ acquires the interpretation as the graviton mass while $G_k$ has the status of a wave-function renormalization. 
Substituting $\Gamma_k^{(hh)}$ into the left-hand side of the Wetterich equation yields
\begin{equation}\label{LHSofTwopoint}
\partial_t \Gamma^{(hh)}_k = \frac{1}{32 \pi }\sqrt{\epsilon_s} \left( (\epsilon^{-1}_s p^2_0 + \vec{p}^{\, 2})\partial_t \frac{1}{G_k} -2 \partial_t \frac{\Lambda_k}{G_k} \right) {\Pi_h}^{ij}_{kl},
\end{equation}
where $\Pi_h$ is the projection tensor which projects a rank two symmetric tensor onto its TT component. The beta function for $\Lambda_k$ can then be read off from the momentum-independent part of the two-point function. In order to obtain the beta function of $G_k$, one can either project onto the $p^2_0$ or $\vec{p}^{\,2}$ component of \eqref{LHSofTwopoint}. We refer to these projections as the $p_0$- and $\vec{p}$-projection, respectively.\footnote{The fact that the two projections yield different RG flows reflects that one is considering the beta functions for different avatars of Newton's coupling which, owed to the Lorentz symmetry breaking of the setup, need to be distinguished on conceptual grounds. Also see \cite{Saueressig:2023tfy} for a more detailed discussion.} The comparison between two projections gives an estimate for the size of Lorentz symmetry breaking effects. 

%-------------------------------------------------------
\section{Beta functions}
\label{sect.4}
%-------------------------------------------------------
The computation of the beta functions encoding the $k$-dependence of the two-point function \eqref{flowtwopoint} can be organized along the lines taken in \cite{Saueressig:2023tfy}. When evaluating the momentum integrals on the rhs, the real part of the parameter $\epsilon_s$ is fixed to $1$ and $-1$, corresponding to Euclidean and Lorentzian signature, respectively. In addition, we retain the $i \epsilon$-prescription for the momentum integration in the Lorentzian setting. The loop integrals are then evaluated in each case. As the main result of this work, this computation shows that the Euclidean and Lorentzian RG flows are identical. This feature follows from the analyticity property of the regulator \eqref{Rchoice} which leads to identical values for the loop integral under the transformation \eqref{redeflap}. As a consequence, there is no need to distinguish between the Euclidean and Lorentzian signature RG flows in the sequel and all our results apply to both cases.

The $k$-dependence of the dimensionless couplings $g_k \equiv G_k k^2$ and $\lambda_k \equiv \Lambda_k /k^2$ is encoded in the beta functions
\begin{equation}\label{betadef2}
\partial_t \lambda_k = \beta_\lambda(g_k,\lambda_k) \, , \qquad \partial_t g_k = \beta_g(g_k,\lambda_k) \, .   
\end{equation}
The explicit computation yields
\begin{equation}\label{betafEHPara}
\begin{split}
	\beta_g = & \, \left(2 + \eta_N \right) g \, , \\
	\beta_\lambda = & \, (\eta_N - 2) \lambda + \frac{g}{4725 \, \lambda ^2 \, \pi} \\ & \bigg(w^1_\lambda +\eta_N \tilde{w}^1_\lambda +  \frac{w^2_\lambda +\eta_N \tilde{w}^2_\lambda}{(1-2 \lambda )^{5/2}}+ \frac{w^3_\lambda  +\eta_N \tilde{w}^3_\lambda}{ (1- \frac{3}{2} \lambda )^{5/2} }\bigg) \, .
\end{split}
\end{equation}
The $w_\lambda^i$ and $\tilde{w}_\lambda^i$ are polynomials in $\lambda$ and listed in the first block of Table \ref{Tab.polypara}.
\begin{table*}[t!]
	\renewcommand{\arraystretch}{1.5}
\begin{center}
\begin{tabular}{ll}
\hline\hline
$\bigg. w^1_\lambda$ & $15525 \lambda ^2+8652 \lambda -2324$\\
\hline
$\bigg. \tilde{w}^1_\lambda$ & $135 \lambda ^2-1236 \lambda +332$\\
\hline
$\bigg. w^2_\lambda$ & $-6 \left( 300545 \lambda ^4 - 527933\lambda ^3+ 383491 \lambda ^2- 127872 \lambda + 16546 \right)$ \\
\hline
$\bigg. \tilde{w}^2_\lambda$ & $2 \left( 180327 \lambda ^4-296517 \lambda ^3+198772 \lambda ^2-60696 \lambda +7178 \right)$ \\
\hline
$\bigg. w^3_\lambda$ & $2 \left(375795 \lambda^4 - 912849 \lambda^3 + 822489 \lambda^2 - 330252 \lambda + 50800      \right)$\\
\hline
$\bigg.  \tilde{w}^3_\lambda$ &$- 150318 \lambda^4 + 344733 \lambda^3  - 288828 \lambda^2 + 105928 \lambda -14688  $\\ 
 \hline  \hline
$\bigg.  w^4_\lambda$ &$-135\lambda^4 + 34776\lambda^3 + 132216\lambda^2 - 433664\lambda + 71120 $\\ \hline
$\bigg. w^5_\lambda$ & $164498565 \lambda ^6-500454234 \lambda ^5+641684601 \lambda ^4 -443566412 \lambda ^3+174167272 \lambda ^2-36783720 \lambda +3259992$ \\
\hline
$\bigg. w^6_\lambda$ & $-4329575145  \lambda ^6+15697970316 \lambda ^5 -23841861804  \lambda ^4+19445619456  \lambda ^3 -8995973376  \lambda ^2+2240897024  \lambda -235003904 $\\
\hline
$\bigg. \tilde{w}^4_\lambda$ & $-\left(2025\lambda^4 + 4968\lambda^3 + 18888\lambda^2 - 61952\lambda + 10160 \right)$ \\
\hline
$\bigg. \tilde{w}^5_\lambda$  & $-98699139 \lambda ^6+285744222 \lambda ^5-347335841 \lambda ^4 +227082460 \lambda ^3-84266056 \lambda ^2+16831752 \lambda  -1413816$ \\
\hline
$\bigg. \tilde{w}^6_\lambda$ & $865915029  \lambda ^6-3011887260   \lambda ^5 +4362687516   \lambda ^4-3374371584   \lambda ^3 +1473101568   \lambda ^2-345042944   \lambda  +33972224  $\\
 \hline \hline
$\bigg. w^7_\lambda$ & $-3699 \lambda ^3+9360 \lambda ^2+8532 \lambda -30016$\\
\hline
$\bigg. w^8_\lambda$ & $-1272480 \lambda ^6+20671730 \lambda ^5-46807002 \lambda ^4 +45990965 \lambda ^3-23327508 \lambda ^2+6039068 \lambda -634560$ \\
\hline
$\bigg. w^9_\lambda$ & $31978800  \lambda ^6-559973655   \lambda ^5+1544967540  \lambda ^4 -1848096396   \lambda ^3+1141774848   \lambda ^2 -361502976   \lambda  +46888960  $ \\
\hline
$\bigg. \tilde{w}^7_\lambda$ \phantom{$(\lambda)$}& $- 4 \left(135 \lambda ^2+780 \lambda -497 \right)$  \\ \hline
$\bigg. \tilde{w}^8_\lambda$  &$-212080 \lambda^4 + 113424 \lambda^3 + 119699 \lambda^2 - 103028 \lambda + 20244$\\ \hline
$\bigg. \tilde{w}^9_\lambda$ &$-394800 \lambda ^3+129321  \lambda ^2 +299328 \lambda -125440$\\
\hline
\hline \hline
\end{tabular}
\caption{\label{Tab.polypara} Polynomials $w^i_\lambda$ and $\tilde{w}^i_\lambda$ introduced in the beta functions \eqref{betafEHPara} (top block), the anomalous dimension \eqref{etaNfEHpara} for the $p_0$-projection (middle block), and $\vec{p}$-projection (lower block).}
\end{center}
\end{table*}
The anomalous dimension can be cast into the form
\begin{equation}\label{etaNfEHpara}
\eta_N = \frac{g \, B_1(\lambda)}{1-g \, B_2(\lambda)} \, . 
\end{equation}
The functions $B_1(\lambda)$ and $B_2(\lambda)$  can be evaluated by projecting onto either the set  $\{ p^2_0 ~\Pi_h, \Pi_h\}$ ($p_0$-projection) or  $\{ {\vec{p}}^{~2} \Pi_h, \Pi_h\}$ ($\vec{p}$-projection). In a Lorentz-covariant setting, these projections would give identical results. Owed to the Lorentz-symmetry breaking contributions, the two projections lead to different results. For the $p_0$-projection we find
 \begin{equation}\label{Bp0projfEHpara}
\begin{split}
	B_1^{p_0}= & \frac{1}{1575 \pi \lambda^4} \Bigg( \frac{w^4_\lambda}{18}
     +\frac{w^5_\lambda}{(1-2 \lambda )^{7/2}} +   \frac{w^6_\lambda}{72  (1- \frac{3}{2} \lambda )^{7/2} }  \Bigg) \, , \\
	B_2^{p_0} = & \frac{1}{4725 \pi \lambda^4} \Bigg( \frac{ \tilde{w}^4_\lambda}{6}+\frac{\tilde{w}^5_\lambda}{(1-2 \lambda )^{7/2}} + \frac{\tilde{w}^6_\lambda}{24  (1- \frac{3}{2} \lambda )^{7/2}} \Bigg) \, ,	
\end{split}
\end{equation}
while the $\vec{p}$-projection leads to
\begin{equation}\label{BpvecprojfEHpara}
\begin{split}
	B_1^{\vec p} = & \frac{1}{315 \pi \lambda^3} \Bigg( \frac{w^7_\lambda}{9} + \frac{w^8_\lambda}{5  (1-2 \lambda )^{7/2} } + \frac{w^9_\lambda}{360  (1-\frac{3}{2} \lambda )^{7/2}}  \Bigg), \\
	B_2^{\vec p} = & \frac{1}{1575 \pi \lambda^2} \Bigg( \frac{\tilde{w}^7_\lambda}{3} + \frac{\tilde{w}^8_\lambda}{(1-2 \lambda )^{5/2}} + \frac{\tilde{w}^9_\lambda}{6 (1- \frac{3}{2} \lambda )^{3/2}} \Bigg) \, . 
\end{split}
\end{equation}
The polynomials $w^i_\lambda$ and $\tilde{w}^i_\lambda$ are given in the second and third block of Table \ref{Tab.polypara}, respectively.

\begin{figure*}
	\includegraphics[width=0.4\textwidth]{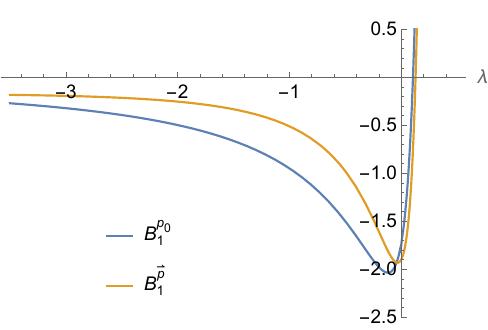} \, 
	\includegraphics[width=0.4\textwidth]{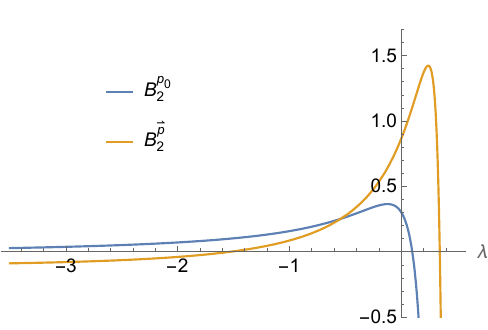}
	\caption{\label{Fig.Bcomp.Para} Plots of the functions $B_1(\lambda)$ (left) and $B_2(\lambda)$ (right) given in Eqs.\ \eqref{Bp0projfEHpara} and \eqref{BpvecprojfEHpara}.}
\end{figure*} 
At this stage, a number of observations are in order. First, the functions $B_1(\lambda)$ and $B_2(\lambda)$ are shown in Fig.\ \ref{Fig.Bcomp.Para}. From these plots, we find that two projection methods give qualitatively similar results. This suggests that even though the spatial momentum regulator breaks Lorentz symmetry, the qualitative properties of the RG flow are projection independent. 

In addition, the analytic structure of $\beta_\lambda$ and $\beta_g$ warrants a detailed discussion. We start with the locus $\lambda = 0$. Fig.\ \ref{Fig.Bcomp.Para} shows that $B_1(\lambda)$ and $B_2(\lambda)$ are regular at $\lambda=0$, despite the poles shown in Eqs.\ \eqref{Bp0projfEHpara} and \eqref{BpvecprojfEHpara}. This feature can also be confirmed analytically by explicitly evaluating the limits of the functions at this point. This property also extends to $\beta_\lambda$. As a consequence, the line $\lambda=0$ is regular with respect to the RG flow for generic values $g$. A brief inspection of the numerators appearing in $\beta_\lambda$, $B_1(\lambda)$, and $B_2(\lambda)$ shows that the beta functions exhibit poles of finite order at
\be\label{gammasing2} 
\begin{split}
\gamma_1^{\rm sing} \; : & \; \left\{ (g,\lambda) \; | \; (\lambda = 1/2, g \not = 0) \right\} \, , \\
\gamma_2^{\rm sing} \; : & \; \left\{ (g,\lambda) \; | \; (\lambda = 2/3, g \not = 0) \right\} \, .
\end{split}
\ee
These lines constitute singularities of the flow equation that cannot be crossed by its solutions. In addition, there is a singular locus caused by the divergence of $\eta_N$. This occurs when the denominator in $\eta_N$ vanishes
\be\label{gammasing}
\gamma^{\rm sing}_3 \; : \; \left\{ (g,\lambda) \; | \; g = 1/B_2(\lambda) \, \right\} \, . 
\ee
Finally, we observe that $\beta_\lambda$, $B_1$, and $B_2$ contain the combination $(1-2\lambda)$ (and also $(1-3/2\lambda)$) with half-integer powers. This leads to branch cuts 
\be\label{gammacut}
\gamma^{\rm cut} \; : \; \left\{ (g,\lambda) \; | \; \lambda > 1/2 \, \right\} \, . 
\ee
In this region, the arguments of the square roots turn negative so that the beta functions take complex values. The comparison with the beta functions computed in \cite{Saueressig:2023tfy} shows that this feature originates from the use of the spatial regulator \eqref{Rchoice}. For a covariant regulator the square roots are absent and there is a well-defined RG flow for $\lambda > 1/2$ as well.

%-------------------------------------------------------
\section{Fixed points and phase diagrams}
\label{sect.4b}
%-------------------------------------------------------
We proceed by constructing the phase diagrams resulting from the beta functions \eqref{betafEHPara}. In addition to the singular lines and branch cuts \eqref{gammasing2}-\eqref{gammacut}, the flow is governed by the interplay of its fixed points $(g_*,\lambda_*)$ where, by definition $\beta_g(g_*, \lambda_*) = \beta_\lambda(g_*, \lambda_*) = 0$. The properties of the flow in the vicinity of such a fixed point are then encoded in the stability matrix
\be
B_j{}^i \equiv \left. \frac{\p \beta_{u^i}}{\partial u^j} \right|_{u = u_*} \, . 
\ee
Defining the stability coefficients $\theta_i$ as minus the eigenvalues of $B$, eigendirections where Re($\theta_i) > 0$ (Re($\theta_i) < 0$) are attracted towards (repelled by) the fixed point as $k \rightarrow \infty$. Hence, UV-attractive eigendirections come with Re($\theta_i) > 0$.

Investigating the fixed points of \eqref{betafEHPara}, we first encounter the Gaussian fixed point (GFP),
\begin{equation}\label{GFP-Parmt}
\text{GFP:} \qquad (g_*,\lambda_*) = (0,0) \, , \quad (\theta_1, \theta_2)= (2, -2).
\end{equation}
The stability properties of this fixed point coincide with canonical power counting, warranting its classification as a Gaussian (or non-interacting) fixed point. The GFP is a saddle point. Its UV-repulsive direction is associated with the coupling $g_k$, so that trajectories with $g_k > 0$ are repelled by this fixed point as $k \rightarrow \infty$. In addition to the GFP, the system also possesses several candidates for non-Gaussian fixed points (NGFPs). These are tabulated in Table \ref{TableNGFPParam}. One finds three candidates 
for the $p_0$-projection while the $\vec{p}$-projection gives rise to two roots only. Both projection schemes support one root situated at a positive Newton's coupling coming with two UV-attractive eigendirections. These are labeled by NGFP$_1$ in Table \ref{TableNGFPParam}. A priori, they come with all properties required for a UV-completion of gravity. The large critical exponents observed in the $\vec{p}$-projection warrant a critical assessment whether these roots correspond to genuine RG fixed points though. The detailed analysis of Appendix \ref{App.C} reveals that the roots seen in the $p_0$- and $\vec{p}$-projection are not connected by a continuous deformation of the beta functions. Thus they do not fall into the same fixed point class. This also provides a natural explanation of the vast difference in the fixed point's position and stability coefficients, as reported in Table \ref{TableNGFPParam}. Moreover, only the NGFP$_1$ seen in the $p_0$-projection is continuously connected to the NGFP found in \cite{Saueressig:2023tfy}. This makes it likely that the NGFP$_1$ seen in the $p_0$-projection constitutes a genuine fixed point while the NGFP$_1$ obtained from the $\vec{p}$-projection is much more susceptible to being a computational artifact instead of a genuine NGFP. Nevertheless, we will construct the phase diagrams for both cases in the sequel. NGFP$_2$ and NGFP$_3$ are located at $g_* < 0$. Hence their flow is disconnected from the physically interested region where Newton's coupling is positive. Thus, these solutions will not play any role in the subsequent discussion.
\begin{table}
{\small
 	\begin{center}
 		\begin{tabular}{ | c | c | c  c  | c  c |  }
 			\hline \hline
 		 Projection & \Big. Fixed Points  &  \multicolumn{2}{c|}{Couplings} & \multicolumn{2}{c|}{Critical Exponents}\\	
 			\hline	
 			~ & ~& $g_*$ & \Big. $\lambda_*$ &  $\theta_1$ & $\theta_2$  \\ \hline \hline
 		\multirow{3}{*}{$p_0$} & \Big. NGFP$_1$ &  $0.89$ &   $-0.40$  &  $5.44$ & $1.25$ \\ 
 		~& \Big. NGFP$_2$ &$-0.17$  & $0.21$            &  \multicolumn{2}{c |}{$1.96 \pm 5.70 i $}\\
 		~& \Big. NGFP$_3$ & $-818.39$  & $-21.94$          & $-31.73$ & $0.33$\\
 			\cline{2-6}  
	\multirow{2}{*}{$\vec{p}$} & \Big. NGFP$_1$ &$301.53$  & $-12.55$  &  $14.11$  &$0.30$\\
	~& \Big. NGFP$_2$ &  $-0.14$  & $0.25$          &  \multicolumn{2}{c |}{$-0.40 \pm 5.61 i $}\\ 
 		\hline \hline
 		\end{tabular}
 	\end{center}
}
 	\caption{ The fixed-point structures for the analytically continued Einstein-Hilbert truncation with $|\epsilon_s|=1$. A fixed point NGFP$_1$, relevant for Asymptotic Safety, is present for all projection schemes. The numerical values of its position and critical exponents differ substantially. This reflects the feature that the NGFP$_1$'s belong to two different branches of fixed points. The solutions NGFP$_2$ and NGFP$_3$ are added for completeness.}\label{TableNGFPParam}
 \end{table}
 
\begin{figure}
 	\centering
 	\includegraphics[width = 0.48\textwidth]{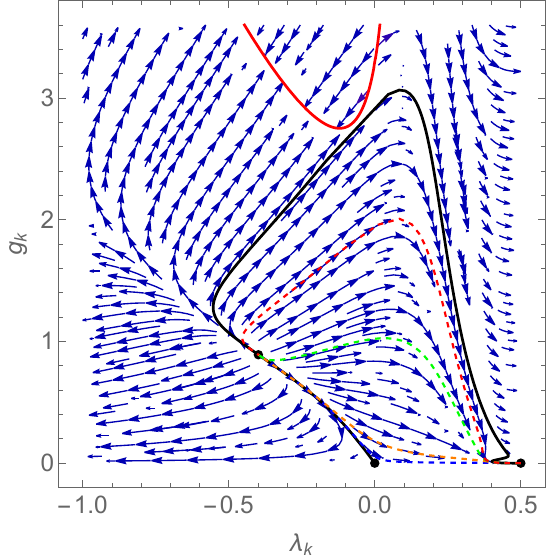} 
 	\caption{\label{FlowDiaFEHParamp0} Phase diagram for the Lorentzian Einstein-Hilbert truncation arising from the beta functions of the $p_0$-projection. 
The arrows point towards the lower values of the coarse-graining scale $k$. We mark the GFP \eqref{GFP-Parmt}, the NGFP$_1$ from Table \ref{TableNGFPParam}, and the IR-FP \eqref{IR-FP-eq} with black dots. The red line denotes the singular locus  $\gamma^{\rm sing}_3$ while the thick black lines label the separatrix and the boundary of the region with RG trajectories connecting NGFP$_1$ to the IR-FP. The colored dashed lines correspond to the RG trajectories along which we analyze the scaling of the couplings. The scaling behavior along these trajectories can be found in the left column of Fig.\ \ref{scalingparmet}. }
 \end{figure}

We proceed by presenting the phase diagrams arising from the $p_0$- and $\vec{p}$-projections. The focus is on the region $g \ge 0$, $\lambda \le 1/2$ which is bounded by $\gamma_2^{\rm sing}$ to its right. The flow in this region is governed by the interplay of the GFP, NGFP$_1$, and the singular locus $\gamma_3^{\rm sing}$. For the $p_0$-projection, the resulting phase diagram is shown in Fig.\ \ref{FlowDiaFEHParamp0}.

The most remarkable feature revealed by the phase diagram is the presence of an IR-FP
\be\label{IR-FP-eq}
\text{IR-FP}: \quad (g_*^{\rm IR}, \lambda_*^{\rm IR}) = (0, 1/2) \, . 
\ee
This fixed point escapes the standard search for fixed points since the beta functions are degenerate for these specific values of the couplings. The presence of the fixed point is then concluded based on scaling properties of RG trajectories in its vicinity. A numerical investigation shows that these indeed follow a power-law scaling
\begin{equation}\label{IR-FP-scaling-Paramt}
\text{IR-FP:} \quad \lambda_*^{\rm IR} - \lambda_k \sim c_1 e^{1.15 t} \, , \; g_k - g_*^{\rm IR} \sim c_2 e^{4 t} \, . 
\end{equation}
Here the constants $c_1$ and $c_2$ specify the RG trajectory and the symbol $\sim$ indicates that Eq.\ \eqref{IR-FP-scaling-Paramt} holds in the vicinity of the IR-FP only. Converting back to the dimensionful graviton mass $\mu_k^2 \equiv -2 \lambda_k \, k^2$ shows that this IR-FP entails a vanishing graviton mass in the limit $k \rightarrow 0$. 

In the next step, we identify two special RG trajectories, the separatrix connecting the NGFP$_1$ to the GFP, and the boundary for the trajectories connecting the NGFP$_1$ to the IR-FP. These are highlighted by the thick black lines. They bound the region in phase space where the trajectories end up in the IR-FP as $k \rightarrow 0$. These trajectories then exhibit a vanishing renormalized graviton mass. In addition to this phase, there is a set of RG trajectories to the left of the separatrix. These either terminate in $\gamma_3^{\rm sing}$ at a finite value of $k$ or reach the endpoint $(g,\lambda) = (0, -\infty)$ at $k=0$. The latter correspond to RG trajectories with a positive renormalized graviton mass.

\begin{figure*}
 	\centering
 	\includegraphics[width = 0.45\textwidth]{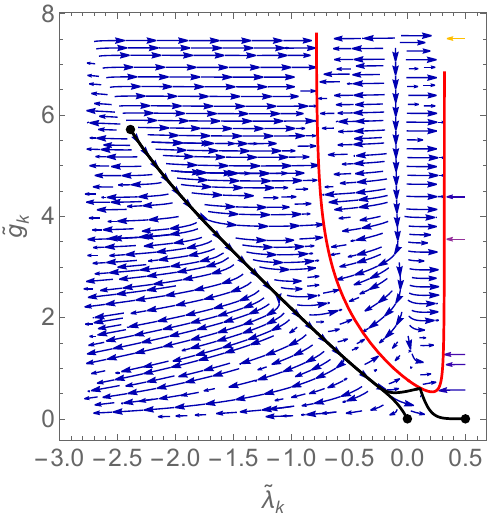} \; \; 
         \includegraphics[width = 0.47\textwidth]{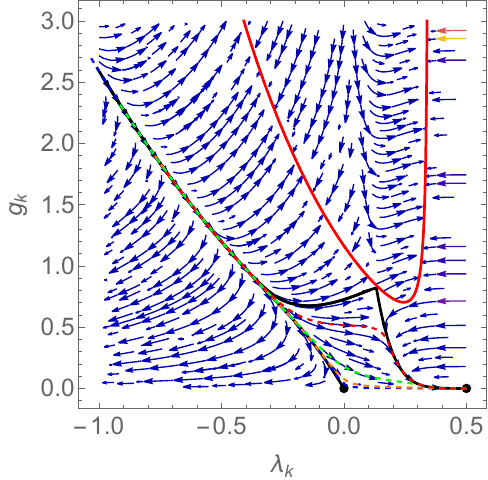}
 	\caption{\label{FlowDiaFEHParam} Phase diagrams for the Lorentzian Einstein-Hilbert truncation based on the $\vec{p}$-projection. The left diagram shows the global properties of the flow, utilizing the map \eqref{redexpcoupl}. The right diagram zooms into the part showing the interplay of the GFP, IR-FP, together with RG trajectories emanating from the NGFP$_1$. The arrows point towards the lower values of the coarse-graining scale $k$. The GFP \eqref{GFP-Parmt}, the NGFP$_1$ from Table \ref{TableNGFPParam}, and the IR-FP are marked by black dots. The red line denotes the singular locus  $\gamma^{\rm sing}_3$ while the thick black lines label the separatrix and the boundary of the region with RG trajectories connecting NGFP$_1$ to the IR-FP. The colored dashed lines correspond to the RG trajectories along which we analyze the scaling of the couplings displayed in the right column of Fig.\ \ref{scalingparmet}.}
 \end{figure*}

The analogous analysis for the $\vec{p}$-projection is summarized in Fig.\ \ref{FlowDiaFEHParam}. 
In this case, the NGFP$_1$ is located at $(g_*, \lambda_*)= (301.53, -12.55)$. In order to be able to capture all relevant features of the phase diagram, we use the  following redefinition of the couplings
\begin{equation}\label{redexpcoupl}
\tilde{g}_k = \text{ln}(g_k+1), \quad \tilde{\lambda}_k = -\text{ln}(\lambda_c-\lambda_k)+ \text{ln} \lambda_c \, . 
\end{equation}
The constant $\lambda_c$ is chosen as $\lambda_c=1/ (2-2~e^{-1/2})$ which ensures that the positions of the GFP and IR-FP remain the same also in the new couplings. The phase diagram for the global flow is then obtained from the beta functions $(\beta_{\tilde{g}}(\tilde{g},\tilde{\lambda}), \beta_{\tilde{\lambda}} (\tilde{g},\tilde{\lambda}))$ and shown in the left panel of Fig.\ \ref{FlowDiaFEHParam}. The right panel uses the original coordinates $(g_k,\lambda_k)$ and zooms into the lower-right corner thereby highlighting the interplay between the GFP, IR-FP and $\gamma_3^{\rm sing}$. Notably, the phase diagram is qualitatively identical to the one shown in Fig.\ \ref{FlowDiaFEHParamp0}. In particular, the IR-FP is again located at \eqref{IR-FP-eq} and induces the characteristic scaling \eqref{IR-FP-scaling-Paramt} for RG trajectories in its vicinity. The new feature is the long funnel connecting the lower-right region to the NGFP$_1$. Essentially, the RG trajectories emanating from NGFP$_1$ and reaching this region are squeezed into an extremely narrow shape which can be almost regarded as a line.

\begin{figure*}
	\includegraphics[width = 0.45\textwidth]{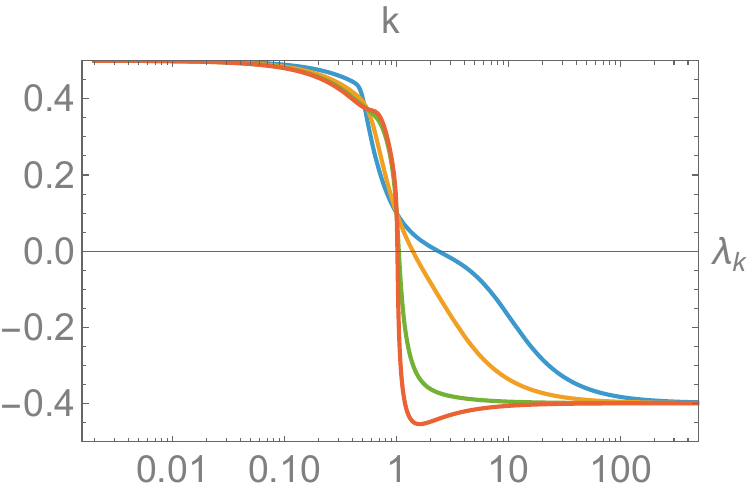} \; \; 
	\includegraphics[width = 0.45\textwidth]{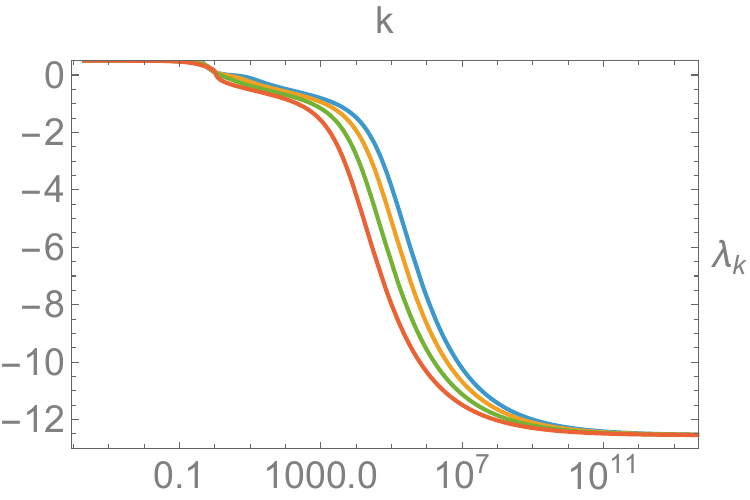} \\[1.4ex]
          \includegraphics[width = 0.45\textwidth]{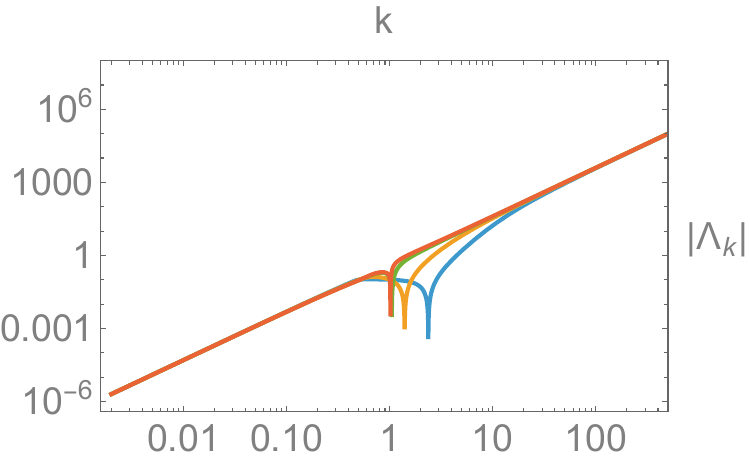} \; \;
	\includegraphics[width = 0.45\textwidth]{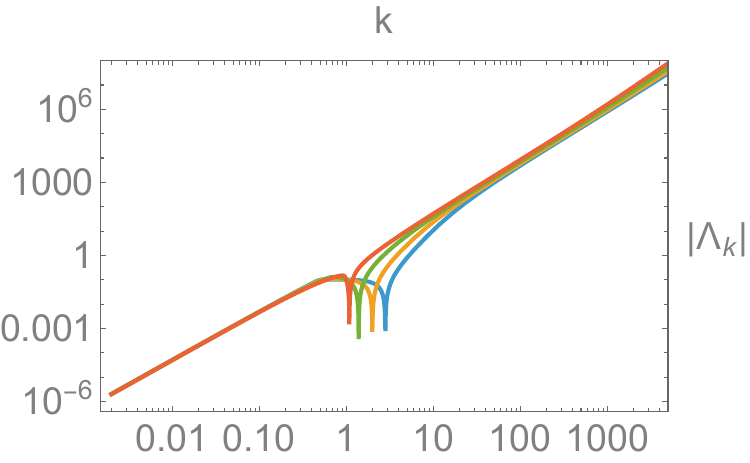} \\
	\caption{\label{scalingparmet} Examples illustrating the $k$-dependence of the dimensionless coupling $\lambda_k$ (top line) and its dimensionful counterpart $\Lambda_k$ (bottom line) for the $p_0$-projection (left column) and $\vec{p}$-projection (right column). The trajectories correspond to the RG trajectories (dashed lines) highlighted in Fig.\ \ref{FlowDiaFEHParamp0}  and Fig.\ \ref{FlowDiaFEHParam} and follow the same color code. For $k \gg 1$ the flow is controlled by the NGFP$_1$ and $\lambda_k$ converges to its fixed point value $\lambda_*$ as $k \rightarrow \infty$.  As $k$ decreases, we find a region near the GFP where the dimensionful coupling $\Lambda_k$ behaves like a constant. In this region one also observes the spikes in $|\Lambda_k|$ which reflect the change of its sign in a logarithmic representation. For $k \rightarrow 0$, the trajectories approach the IR-FP. The feature that the IR-FP drives the dimensionful $\Lambda_k$ to zero is then readily seen from the diagrams in the bottom row. In all plots, $k$ is measured in Planck-units so that the crossover between the non-Gaussian and Gaussian fixed point occurs at $k \approx 1$. 
}
\end{figure*}

We complete the discussion of the phase diagrams by displaying the dependence of $\lambda_k$ on the coarse graining scale $k$ (measured in Planck units)  along the RG trajectories highlighted in Figs.\ \ref{FlowDiaFEHParamp0} and \ref{FlowDiaFEHParam}. The result is shown in Fig.\ \ref{scalingparmet}. All curves interpolate between the NGFP$_1$ in the limit $k \rightarrow \infty$ and the IR-FP as $k \rightarrow 0$. The logarithmic plots showing the $k$-dependence of the dimensionful coupling $\Lambda_k$ establish that the IR-FP implies $\lim_{k \rightarrow 0} \Lambda_k = 0$. The scaling laws \eqref{IR-FP-scaling-Paramt} can be verified by tracking the convergence of $\lambda_k$ towards $\lambda_*^{\rm IR}$ using a logarithmic scale for the difference.

We close this section with an important remark. We recall \cite{Bonanno:2020bil,Saueressig:2023irs} that predictions based on the asymptotic safety mechanism should be based on renormalized couplings. These are obtained at the endpoint of an RG trajectory in the limit $k=0$ where all quantum corrections have been taken into account. Applying this philosophy to the phase diagrams obtained from the $p_0$-projection and the $\vec{p}$-projection reveals an astonishing feature: despite the apparent differences in Figs.\ \ref{FlowDiaFEHParamp0} and \ref{FlowDiaFEHParam} their predictions are actually identical: the values for $\Lambda_k$ compatible with Asymptotic Safety provided by these NGFP are
\be\label{key-result}
\Lambda_0 \le 0 \, , 
\ee
with $\Lambda_0 = 0$ related to the IR-FP attractor. In other words, the squared mass appearing in the Lorentzian graviton propagator given in the first line of Table \ref{table.hessian.parame} may either be positive or zero. The latter case arises from an attractor mechanism without the need of fine-tuning initial conditions to a fixed value. We find it remarkable that both Lorentzian and Euclidean signature computations carried out in this work agree on this feature.

%-------------------------------------------------------
\section{Summary and Outlook}
\label{sect.5}
%-------------------------------------------------------
The transition from Euclidean to Lorentzian signature is a key step in developing the gravitational asymptotic safety program. In comparison to the Euclidean setting, Lorentzian signature computations require an additional geometric structure: a preferred direction which serves as time. Our work implements this structure via the ADM decomposition of the metric field. Our focus is then on the connection of the Euclidean and Lorentzian settings through the analytic continuation of the lapse function \eqref{redeflap}.  This setting is used to determine the flow of the graviton two-point function with the propagator and interactions extracted from the gauge-fixed Einstein-Hilbert action. We explicitly establish that the Lorentzian two-point function resulting from the analytic continuation has the causal structure of the Feynman propagator. Moreover, it is shown that the beta functions obtained from the Lorentzian and Euclidean signature computations are identical. A key ingredient underlying these results is the choice of a spatial regulator \eqref{Rchoice} which renders loop-computations finite without introducing new poles in the complex momentum plane. These properties come at the expense that the regulator breaks Lorentz covariance explicitly. Notably, this is the first time that a two-point function of an ADM field has been computed in the Lorentzian signature setting.

The phase diagrams determining the flow of the graviton mass with respect to the coarse-graining scale are given in Figs.\ \ref{FlowDiaFEHParamp0} and \ref{FlowDiaFEHParam}. Since the ADM decomposition allows to distinguish between spatial and time-components of the external momentum, the flow diagram can be constructed by reading off the wave function renormalization from the time-part ($p_0$-projection) or the spatial components ($\vec{p}$-projection) of the external four-momentum. Both projections give rise to a non-Gaussian renormalization group fixed point suitable for Asymptotic Safety. The analysis of Appendix \ref{App.C} reveals that these fixed points belong to different families which explains their substantial difference in position and stability coefficients. Moreover, a direct comparison with the fixed point results reported for a covariant regulator \cite{Saueressig:2023tfy} shows that the $p_0$-projection fixed point is continuously connected to the one found in the covariant setting. Our analysis based on interpolating parameters, given in Appendix \ref{App.C}, suggests that the NGFP$_1$ seen in the $p_0$-projection is fairly robust. 

The NGFP$_1$ reported for the $\vec{p}$-projection does not belong to this network of NGFPs though. While this provides a natural explanation for the large critical exponents reported in Table \ref{TableNGFPParam}, this also makes this root much more prone to be an artifact of the approximation. Establishing (or refuting) this root as a genuine fixed point will require a refined truncation. This could be done by including the gravitational form factor which characterizes the full momentum dependence of the two-point function. We will implement this improvement in our future work. As a corrollary, we note that covariant regulators seem to lead to beta functions which are more robust in terms of deformations than the analogous result obtained from the spatial regulator \eqref{reg-spatial}.

Remarkably, the phase diagrams are qualitatively similar and agree on their predictions for the graviton mass compatible with Asymptotic Safety. Moreover, the phase diagrams capturing the flow at Lorentzian signature agree with the ones found for the ADM decomposition \cite{Saueressig:2023tfy} and the covariant setting \cite{Christiansen:2014raa}, carried out in a Euclidean background. In particular, the IR-attractor driving the graviton mass to zero dynamically operates in all cases. 

It would be interesting to extend the results reported in this work in various ways. First, one could extend the momentum dependence of the graviton two-point function tracked in our work by promoting the $p^2$-term to a form factor $f(p^2)$,  capturing the full momentum dependence along the lines \cite{Bosma:2019aiu,Knorr:2019atm,Knorr:2022dsx,Knorr:2021iwv,Bonanno:2021squ,Fehre:2021eob}. The flow equation should then determine the analytic structure of the propagator including potential poles in the complex momentum plane. Equivalence of the Euclidean and Lorentzian setting at the level of form factors then requires that this generalization does not lead to new poles in the complex momentum plane with are crossed when performing the analytic continuation of the lapse function. This property is highly non-trivial and should be established based on explicit computations.

Second, it would be interesting to find a Lorentz-covariant way to carry out computations within the ADM formalism. Notably, the spatial momentum regulator is not the only ingredient giving rise to Lorentz symmetry breaking contributions to the RG flow \cite{Saueressig:2023tfy}. At this stage, the construction of a Lorentz-covariant flow on a foliated spacetime, using a covariant regulator,  would be desirable.  Such a construction requires a non-linear field redefinition of the ADM fields \cite{Korver:2024sam}. Similarly to the definition of the fluctuation field based on the linear and exponential split in covariant computations \cite{Ohta:2016npm}, such a redefinition of the fluctuation field may lead to new properties in the RG flow that signal that ultimately, one is quantizing a different theory. We will return to this point in future work \cite{WangInPrep2}.

Finally, the present computation is readily generalized to other two-point functions and more general backgrounds. In this context, the two-point function of the trace mode may be of special interest since this may be related to correlation functions measured within the Causal Dynamical Triangulations program \cite{Ambjorn:2008wc, Ambjorn:2016fbd} along the lines \cite{Knorr:2018kog}. Considering that the trace correlator may depend on the gauge condition, one may need the method developed in \cite{Igarashi:2019gkm} combining the Quantum Master Equation and Wilsonian RG flow equation to extract the BRST-invariant flow. Moreover, one should investigate to which extent the results for the correlation functions depend on the choice of foliation. Along this line, the results recently obtained in \cite{Banerjee:2024tap}, showing that such a dependence vanishes at the level of the complexified heat-kernel diagonal, are very encouraging. This will also be considered in future work.

%-------------------------------------------------------
\section*{Acknowledgments}
We thank J.\ Ambj{\o}rn, T.\ Borck, T.\ Budd, M.\ Becker, R.\ Loll, D.\ Nemeth, and C.\ Wetterich for discussion. Moreover, we thank R.\ Banerjee and M.\ Niedermaier for their insightful comments on the initial manuscript. We also thank the referee for their constructive comments and suggestions on our work. JW acknowledges the China Scholarship Council (CSC) for
financial support.
%-------------------------------------------------------
\section*{Data availability}
The data that support the findings of this article are openly available \cite{WangCodes}.
%-------------------------------------------------------
\appendix
\section{The ghost sector}
\label{App.B}
%-------------------------------------------------------
The ghost term appearing in Eq.\ \eqref{actionansatz} is conveniently split into a scalar and vector part
\be 
\Gamma^{\text{ghost}} = \Gamma_{\text{scalar}}^{\text{ghost}} + \Gamma_{\text{vec}}^{\text{ghost}} \, . 
\ee
The Faddeev-Popov procedure yields
\begin{equation}\label{Vghost}
  	\begin{split}
  		\Gamma_{\text{scalar}}^{\text{ghost}}  =& \int d\tau d^3 y   \, \sqrt{\epsilon_s}\,\bar{c} \,[ \partial_\tau^2 c N +  \partial_\tau b^k \partial_k N - \partial_\tau N N_j \sigma^{ij}\partial_i c \\
&+  \partial^i \partial_\tau N_i c +  \partial^i b^k \partial_k N_i +   \partial^i N_k \partial_i b^k +  \partial^i \sigma_{ki} \partial_\tau b^k \\
&- \partial_\tau \sigma_{ik}\partial^i b^k+  \partial^i \sigma^{kl} N_k N_l \partial_i c + \epsilon_s \partial^i N^2 \partial_i c \\
& - \frac{1}{{2}} ~ \partial_\tau c \partial_\tau  \bar{\sigma}^{ij}\sigma_{ij}- \frac{1}{{2}} \partial_\tau b^k \partial_k \bar{ \sigma}^{ij}\sigma_{ij}-  \partial_\tau N^i \partial_i c],
  	\end{split}
  \end{equation}
and
\pagebreak[4]
 \begin{equation}\label{Sghost}
  	\begin{split}
  		\Gamma_{\text{vec}}^{\text{ghost}}  =& \int d\tau d^3y  \, \sqrt{\epsilon_s} \, \bar{b}^i \, \Big[\frac{1}{\epsilon_s}\partial^2_\tau N_i c + \frac{1}{\epsilon_s}\partial_\tau \sigma^{kl} N_k N_l \partial_i c \\
& +  \partial_\tau N^2 \partial_i c  -   \partial_i \partial_\tau N c 
  		- \partial_i N_m \partial^m c \\
&+ \partial_i N N_k \sigma^{kl} \partial_l c +   \partial^j N_i \partial_j c  +   \partial^j N_j \partial_i c  \\
&
  		 - \frac{1}{2}   \partial_i  c \partial_\tau \delta^{mn}\sigma_{mn}+    \partial^j c \partial_\tau \sigma_{ij} 
  		+\frac{1}{\epsilon_s}  \partial_\tau b^k \partial_k N_i \\
&+ \frac{1}{\epsilon_s}  \partial_\tau N_k \partial_i b^k +  \frac{1}{\epsilon_s}  \partial_\tau \sigma_{ki} \partial_\tau b^k  -   \partial_i b^k \partial_k N \\
  		&- \frac{1}{2}   \partial_i  b^k \partial_k \delta^{mn}\sigma_{mn}-   \partial_i \sigma_{jk} \partial^j b^k  +    \partial^j b^k \partial_k \sigma_{ij} \\
&+    \partial^j \sigma_{jk} \partial_i b^k+   \partial^j \sigma_{ik} \partial_j b^k \Big].
  	\end{split}
  \end{equation}
Here all derivatives act on the right, for instance, $\partial_\tau N_i c =N_i (\partial_\tau c) +(\partial_\tau N_i) c$. In addition, all indices are raised and lowered with the flat background metric $\delta_{ij}$.
%------------------------------------------------
\section{The web of fixed points}
\label{App.C}
%-------------------------------------------------------
At this stage, it is also important to understand whether and how the NGFPs reported in Table \ref{TableNGFPParam} are related. In addition, it is interesting to establish their connection with the NGFPs found with a covariant regulator in the Euclidean setting \cite{Saueressig:2023tfy}. In this appendix, we establish these connections by resorting to interpolation schemes for the momentum projection and regulators.
\begin{figure*}[t!]
 \includegraphics[width=0.45\textwidth]{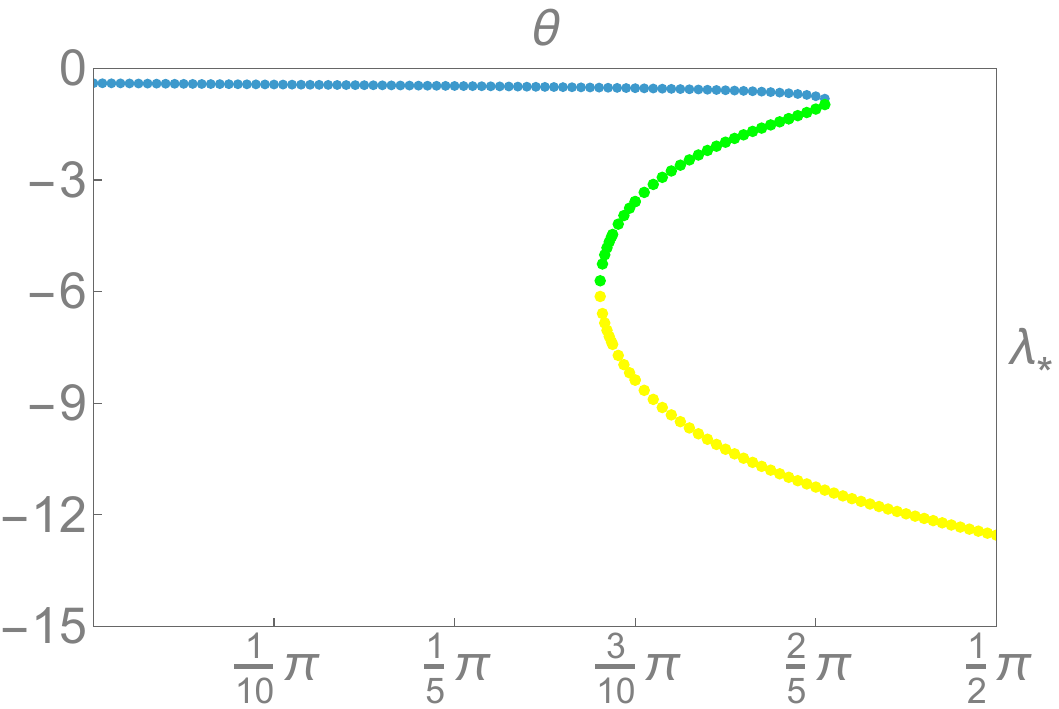}
\includegraphics[width=0.45\textwidth]{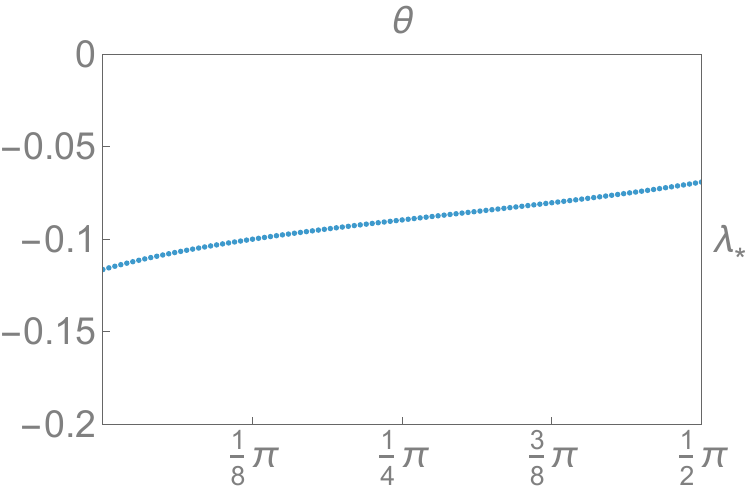}
\caption{\label{Fig.interpolation} Interpolation between the NGFP$_1$ seen in the $p_0$- and $\vec{p}$-projections based on the one-parameter family of projections \eqref{projection-space-int}. The analysis for the spatial regulator \eqref{Rchoice} is shown in the left panel and establishes that the NGFP$_1$ in Table \ref{TableNGFPParam} are not connected by a continuous deformation. The right panel establishes that for the NGFPs reported in \cite{Saueressig:2023tfy}, the converse result holds: in this case the fixed points do belong to the same family.}
\end{figure*}

We start by projecting the RG flow onto the following one-parameter generalization of eq.\ \eqref{projection-space}
\be\label{projection-space-int}
\begin{split}
\Gamma^{(hh)}_k = & \frac{\sqrt{\epsilon_s}}{32 \pi G_k} \left( (\epsilon^{-1}_s p^2_0 \cos \theta + \vec{p}^{\, 2} \sin\theta) -2 \Lambda_k \right) {\Pi_h}^{ij}_{kl} \, .
\end{split}
\ee
The parameter $\theta$ is independent of $k$ and takes values on the interval $[0, \pi/2]$. The ansatz \eqref{projection-space-int} interpolates between the $p_0$-projection for $\theta=0$ and the $\vec{p}$-projection for $\theta=\pi/2$. The beta functions resulting from this projection allow to trace the fixed-point structure related to the NGFP$_1$ as a function of $\theta$. The result is shown in the left panel of Fig.\ \ref{Fig.interpolation}. The analysis shows that the NGFP$_1$ visible in the $p_0$-projection belongs to a continuous family of fixed points which exists on the interval $\theta \in [0, 2/5\pi]$. One also finds that at $\theta \approx 0.3 \pi$, a new pair of fixed points emerges from the complex plane. At $\theta \approx 2/5\pi$ one of the new fixed points collides with the family of NGPF$_1$ from the $p_0$-projection and both fixed points vanish into the complex plane. The NGFP$_1$ seen in the $\vec{p}$-projection then corresponds to the remaining fixed point of the new pair. Hence the NGFP$_1$ seen in the $p_0$- and $\vec{p}$-projections are not connected by a continuous deformation in $\theta$ and belong to different fixed point families. This also explains the substantial difference in the critical exponents reported in Table \ref{TableNGFPParam}.

Notably, the fixed point annihilation seen in the left panel of Fig.\ \ref{Fig.interpolation} is specific to the use of the spatial regulator used in the present work. Applying the interpolation \eqref{projection-space-int} to the covariant computation \cite{Saueressig:2023tfy}, which uses a Lorentz-covariant regulator at Euclidean signature, one finds that the NGFPs obtained from the $p_0$- and $\vec{p}$-projections are related by a continuous deformation in the parameter $\theta$. This feature is depicted in the right panel of Fig.\ \ref{Fig.interpolation}.

\begin{figure}
 \includegraphics[width=0.45\textwidth]{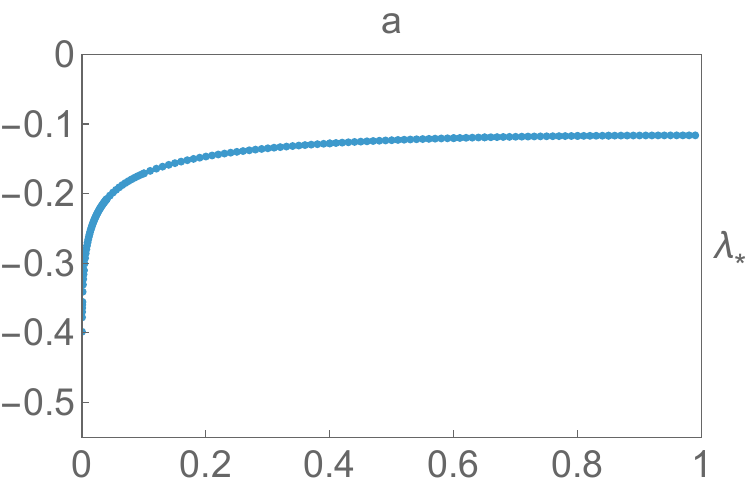}
 \caption{\label{Fig.int.a}{The position of the NGFP$_1$ from the $p_0$-projection as a function of the parameter $a$ introduced in the one-parameter family of regulator functions \eqref{Reg-gen}. The curve interpolates continuously between the NGFP$_1$ given in Table \ref{TableNGFPParam} for $a=0$ and the NGFP$_1$ obtained for the covariant regulator \cite{Saueressig:2023tfy} for $a=1$.}
}
\end{figure}

Finally, we clarify the connection between the NGFPs found when using spatial and covariant regulators in the Euclidean signature setting. For this purpose, we generalize the spatial regulator \eqref{reg-spatial} to the one-parameter family of regulators
\be\label{Reg-gen}
R_k(p_0^2,\vec{p}^{\,2}) = \left(k^2 - a p_0^2 - \vec{p}^{\,2}\right) \Theta\left(k^2 - a p_0^2 - \vec{p}^{\,2}\right) \, . 
\ee
The parameter $a \in [0,1]$ and \eqref{Reg-gen} interpolates between the spatial regulator employed in the main part of this work for $a=0$ and the Lorentz-covariant regulator employed in \cite{Saueressig:2023tfy} for $a=1$. We then trace the dependence of the NGFP$_1$ as a function of $a$. For the sake of conciseness, we limit the discussion to the $p_0$-projection. The position of $\lambda_*$ for the NGPF$_1$ is then shown in Fig.\ \ref{Fig.int.a}.
On this basis, we conclude that the NGFP$_1$ seen in the $p_0$-projection found in this work and in \cite{Saueressig:2023tfy} are indeed connected by a continuous deformation and correspond to the same fixed point. In combination with the results of Fig.\ \ref{Fig.interpolation} this implies that this result does not extend to the $\vec{p}$-projection though.

%----------------------------------------------------------
%\newpage

%----------------------------------------------------------
%----------------------------------------------------------

%----------------------------------------------------------

%----------------------------------------------------------
\end{document}